\def\CVS $#1: #2 #3${\gdef\theCVSId{\url{#3}}}%
\CVS $Id: final-decor2004.tex,v 1.1 2004/09/07 15:36:47 slacour Exp $

\documentclass[english]{article-hermes}

\usepackage{color}
\usepackage{xspace}
\usepackage[latin1]{inputenc}
\usepackage[T1]{fontenc}
\usepackage[obeyspaces]{url}
\usepackage[dvips]{graphicx}
\usepackage{alltt}

\newcommand{\java}{\textsc{Java}\xspace}
\newcommand{\corba}{\textsc{Corba}\xspace}
\newcommand{\ccm}{\textsc{ccm}\xspace}
\newcommand{\etc}{\emph{etc.}\xspace}

\begin{document}

\title[Deployment of components on a grid]{A Software Architecture \\
for Automatic Deployment \\
of CORBA Components \\
Using Grid Technologies}

\author{S\'ebastien Lacour \andauthor Christian P\'erez \andauthor
Thierry Priol}

\address{IRISA~/~INRIA \\
Campus de Beaulieu \\
35042 Rennes, France \\[3pt]
\texttt{\{Sebastien.Lacour,Christian.Perez,Thierry.Priol\}@irisa.fr}}

\resume{%
Les composants logiciels sont une solution bien adapt\'ee pour
construire des applications complexes de calcul scientifique
destin\'ees \`a \^etre ex\'ecut\'ees sur une grille de calcul.
Cependant, le d\'eploiement d'applications complexes \`a base de
composants sur une grille est une t\^ache particuli\`erement ardue.
Pour \'eviter d'avoir \`a faire face directement au grand nombre
d'ordinateurs de la grille et \`a leur h\'et\'erog\'en\'eit\'e, la
phase de d\'eploiement d'application doit \^etre automatis\'ee.
Cet article d\'ecrit une architecture de d\'eploiement automatique
d'applications \`a base de composants sur grille de calcul.  En
partant du mod\`ele de composants \corba (\ccm), ce papier d\'etaille
les \'etapes du d\'eploiement de composants et les acteurs en
pr\'esence: un intergiciel d'acc\`es aux ressources de la grille
(\`a l'instar de OGSI), un mod\`ele de d\'eploiement de composants,
une description \'etendue de l'application et un planificateur de
d\'eploiement.}

\abstract{%
Software components turn out to be a convenient model to build complex
applications for scientific computing and to run them on a
computational grid.  However, deploying complex, component-based
applications in a grid environment is particularly arduous.  To
prevent the user from directly dealing with a large number of
execution hosts and their heterogeneity within a grid, the application
deployment phase must be as automatic as possible.
This paper describes an architecture for automatic deployment of
component-based applications on computational grids.  In the context
of the \corba Component Model (\ccm), this paper details all the steps
to achieve an automatic deployment of components as well as the
entities involved: a grid access middleware and its grid information
service (like OGSI), a component deployment model, as specified by
\ccm, an enriched application description and a deployment planner in
order to select resources and map components onto computers.}

\motscles{Composants \corba, d\'eploiement automatique, grilles de
calcul.}



\proceedings{DECOR'04, Déploiement et (Re)Configuration de Logiciels}{187}

\maketitlepage


\section{Introduction}

Modern software development approaches are often suspected of not
providing the level of performance which high-end parallel computers
would offer. However, new scientific applications become more and more
complex, involving several simulation codes coupled together to obtain
more accurate simulations.
For example, multi-physics simulations aim to simulate various
physics, each of them implemented by a dedicated code, to increase the
accuracy of simulation. It is becoming clear that a radical shift in
software development should occur to handle the increasing complexity
of such applications. Moreover, the computing infrastructure should
provide the level of performance to running such applications within a
reasonable time frame. A computational grid is by no doubt a computing
infrastructure that could deliver this level of performance.  It is a
set of high-performance computing resources connected to the Internet
and managed by a middleware that gives transparent access to resources
wherever should they be located in the network.

Software components turn out to be a convenient model to build
multi-physics applications for scientific computing and to run them on
a computational grid~\cite{PerPriRib03IJHPCA, CCA1}. Each simulation
code can be encapsulated into a component.
Such an approach raises several difficult problems such as
encapsulation of parallel simulation codes into software components
and efficient communication between components in the presence of
various high-performance networking technologies. We already proposed
solutions \cite{PerPriRib03IJHPCA, DenPerPri03FGCS} to those problems
in the context of the \corba component model (\ccm)~\cite{CCM}. For
better acceptance of component-based applications running on a grid,
the deployment phase should be as automatic as possible while taking
into account application constraints (memory, \etc) and/or user
constraints. While environments like ProActive~\cite{proactive} are
able to deal with Grid middleware, they do not support application
and/or user constraints: the mapping of virtual nodes to physical
nodes has to be provided manually and network constraints seem difficult to
handle.

This paper presents an architecture for automatic deployment of
component-based applications on computational grids. 
Section~\ref{sec:architecture} details all the necessary entities and
their relationships to achieve an automatic deployment of components.
Examples of these entities are presented with respect to the prototype
we are currently developing. Before the conclusion,
Section~\ref{sec:challenges} gives an overview of the upcoming
challenges.


\section{Architecture for Automatic Deployment of Components}
\label{sec:architecture}

The \corba component model contains a deployment model that specifies
how a particular component can be installed, configured and launched
on a machine. The specifications do not deal with the problem of
selecting machines and rely on a ServerActivator daemon to actually
launch component servers.

The proposed architecture aims to describe the entities needed for an
automatic deployment as well as their relationships. These entities
can be grouped into three parts, each of them actually corresponding
to a phase of the deployment process: the inputs (the component
assembly and a grid resource description), the planner, which selects
the resources and maps each component on a computer, and the actual
deployment of the components on the selected resources. This section
reviews these entities and mentions a few issues which have already
been tackled within our prototype.


\subsection{Information Description}
\label{sub:information}

Two pieces of information are required for automatic deployment: a
description of the component-based application to deploy and a
description of the grid resources on which the application may be
deployed.

\subsubsection{Component-Based Application Description}
\label{subsub:appli}

Within the context of the \corba Component Model (\ccm,~\cite{CCM}),
an application is made of a set of components, called a component
assembly package. It is an archive provided by the user to the
deployment tool.  It includes, among other files, the assembly
description which describes all the components of the assembly and
their interconnections.

The assembly and component descriptors can express various
requirements such as the processor architecture and the operating
system required by a component implementation. A component may have
environmental or other dependencies, like libraries, executables,
\java classes, \etc Another possible requirement is component
collocation: components may be free or partitioned to
a single process or a single host, meaning that a group of component
instances will have to be deployed in the same process or on the same
compute node.

\subsubsection{Grid Resources}
\label{subsub:grid_resources}

Information about grid resources includes not only compute and storage
resource information, but also network description. While compute and
storage resource description is rather well mastered (computer
architecture, number and speed of CPUs, operating system, memory size,
storage capacity, \etc), network description received less attention.
We have proposed~\cite{LacPerPri2004RR5221} a scalable model for grid
network topology description and have implemented it on top of
MDS2~\cite{CzaFitFosKes2001HPDC}, the information service of the Globus
Toolkit~\cite{GlobusWeb} version~2.

The deployment tool requires a pointer to a resource information
service to be able to automatically find adequate resources.
Depending on the type of resource information service, the pointer can
be a path to a local file, a URL, or a distinguished name (DN), host
and port to access an LDAP tree.  Our
prototype~\cite{LacPerPri2004RR5221} currently supports local file
access, HTTP(S) and (GSI)FTP protocols as well as LDAP~/~MDS2 query.


\subsection{Deployment Planning}
\label{sub:planning}

The deployment planner is responsible for 1)~selecting the computers
which will run the components and the component servers, 2)~selecting
the network links (or network technology) to interconnect the
components, and 3)~mapping the component servers onto the selected computers.
The input of the deployment planning algorithm is made of the
application description and the resource description, as explained in
Subsection~\ref{sub:information}.

The output of the deployment planner is a deployment plan that
describes the mapping of the components onto component servers and the
mapping of these component servers onto the computers of the grid.
The deployment plan should also specify 1)~in what order processes
must be launched by the deployment tool, 2)~how data must flow from
the output of certain processes to the input of other processes,
3)~what network connections must be established between every pair of
processes. For instance, items~1) and~2) are necessary for \corba
applications, where a Naming Service needs to be launched, and its
reference needs to be passed to the other processes.

Our prototype is currently based on a simple round-robin deployment
planning algorithm. It is just  a proof of concept.


\subsection{Actually Launching Components on a Grid}

Once the deployment plan has been obtained from the previous step, the
component-based application is launched and configured according to
the \corba component model. The technical point is that the selected
machines are assumed not to contain any component activator or
component server. That is why a job submission method is
needed to launch this very first process. This step is fully
compatible with the \ccm deployment model~\cite{LacPerPri2004CD} but
needs more work to comply with the MDA deployment
specification~\cite{MDA-deployment}.
The deployment tool manages two sorts of handles: \corba references
and handles returned by the grid access middleware.  Both are useful
to control application processes, like cancel, suspend, or restart
their execution.

To face the diversity of grids, the deployment tool should support
various grid access middleware such as the Globus
Toolkit~\cite{CzaFosKarKesMarSmiTue1998IPPS}, OSGA,
Condor~\cite{CondorG}, \etc Our prototype illustrates how \corba
components can be deployed on a computational grid using the Globus
Toolkit~\cite{GlobusWeb}: more details are provided
in~\cite{LacPerPri2004CD}.


\section{Efficient Automatic Deployment}
\label{sec:challenges}

While a few issues have already been addressed as mentioned in
Section~\ref{sec:architecture}, there remains a number of issues to
achieve an efficient automatic deployment which are mainly related to
the constraints attached to an application. The central issue is to
understand what a user expects from an automatic deployment tool and
what is possible, like prediction of the behavior of a
component~\cite{FurMayGouNewFieDar2002JPC} for example.


\subsection{Enriching the Application Description}

The constraints attached to a component assembly package mainly focus
on enabling the execution of the component. New kinds of constraints
could be useful, like communication requirements (latency, bandwidth,
\etc) or a description of the behavior of the application with respect
to specific resources~\cite{FurMayGouNewFieDar2002JPC}. \ccm
specifications allow new constraints to be added to the component
assembly package as in~\cite{WanRodGil2003} for example. A major issue
is to standardize useful constraints.


\subsection{Taking User-Level Constraints into Account}

The deployment planning algorithm (see Subsection~\ref{sub:planning})
requires a \emph{goal}~\cite{Sekitei} to produce a deployment plan.
For example, do we want to minimize the execution time or do we want
the application to run at a particular site, close to a visualization
node? Those constraints are not specific to the application itself,
they are user-level constraints. They belong neither to the
application description nor to the grid resource description. To take them into
account, a third kind of information needs to be defined.


\subsection{Deployment Planning Algorithm}

The deployment planning algorithm of our prototype is too simple to
satisfy the constraints mentioned above..  More sophisticated
algorithms like
Sekitei~\cite{Sekitei} exist, but the question is to determine if they
are suitable for our purpose. To support a variety of
application-level and user-level constraints, the planner needs to be
very customizable. Do we need a general purpose deployment planning
algorithm? Or do we need a collection of specialized algorithms?  The
latter solution may give better results, since we can imagine that an
application may provide its own fine-tuned deployment algorithm.


\section{Conclusion}

On the one hand, software component technologies appear to be a
convenient model to handle the complexity of multi-physics
simulations. On the other hand, grids promise to offer the necessary
level of performance for such applications. This paper has presented
an architecture to achieve automatic deployment of component-based
applications in a grid environment. The central entity is the
deployment planner which has to select resources and map components on
them to achieve a goal. The planner requires the description of both
the component assembly and grid resources. It generates a deployment
plan which controls the \ccm deployment with the help of a job
submission method. Remaining issues of our ongoing work include the
definition of useful application-level constraints, management of
user-level constraints, and integration of efficient deployment
planning algorithms.


\bibliography{decor2004}

\begin{thebibliography}{}

\bibitem[ARM~99]{CCA1}
\textsc{Armstrong R.}, \textsc{Gannon D.}, \textsc{Geist A.}, \textsc{Keahey
  K.}, \textsc{Kohn S.}, \textsc{McInnes L.}, \textsc{Parker
  S.}\andname{}\textsc{Smolinski B.}, \guilo{}Toward a Common Component
  Architecture for High-Performance Scientific Computing\guilf{},
\newblock \Inname{} \textit{Proc. of the 8th Intl. Symp. on High Performance
  Distributed Computation}, \Aug{} 1999.

\bibitem[BAU~02]{proactive}
\textsc{Baude F.}, \textsc{Caromel D.}, \textsc{Mestre L.}, \textsc{Huet
  F.}\andname{}\textsc{Vayssi{\`e}re J.}, \guilo{}Interactive and
  Descriptor-based Deployment of Object-Oriented Grid Applications\guilf{},
\newblock \Inname{} \textit{Proc. of the 11th Intl. Symp. on High Performance
  Distributed Computing}, July 2002,  \pagesname{} 93-102.

\bibitem[CZA~98]{CzaFosKarKesMarSmiTue1998IPPS}
\textsc{Czajkowski K.}, \textsc{Foster I.}, \textsc{Karonis N.},
  \textsc{Kesselman C.}, \textsc{Martin S.}, \textsc{Smith
  W.}\andname{}\textsc{Tuecke S.}, \guilo{}A Resource Management Architecture
  for Metacomputing Systems\guilf{},
\newblock \Inname{} \textit{Proc. of the Workshop on Job Scheduling Strategies
  for Parallel Processing}, \volumename{} 1459 \ofname{} \textit{LNCS}, 1998,
  \pagesname{} 62-82.

\bibitem[CZA~01]{CzaFitFosKes2001HPDC}
\textsc{Czajkowski K.}, \textsc{Fitzgerald S.}, \textsc{Foster
  I.}\andname{}\textsc{Kesselman C.}, \guilo{}Grid Information Services for
  Distributed Resource Sharing\guilf{},
\newblock \Inname{} \textit{Proc. of the 10th IEEE Intl. Symp. on
  High-Performance Distributed Computing}, San Francisco, CA, \Aug{} 2001,
  \pagesname{} 181-194.

\bibitem[DEN~03]{DenPerPri03FGCS}
\textsc{Denis A.}, \textsc{Pérez C.}\andname{}\textsc{Priol T.},
  \guilo{}{PadicoTM}: An Open Integration Framework for Communication
  Middleware and Runtimes\guilf{},
\newblock \textit{FGCS}, \volumename\ 19, 2003,  \pagesname{} 575-585.

\bibitem[FRE~01]{CondorG}
\textsc{Frey J.}, \textsc{Tannenbaum T.}, \textsc{Livny M.}, \textsc{Foster
  I.}\andname{}\textsc{Tuecke S.}, \guilo{}{Condor-G}: A Computation Management
  Agent for Multi-Institutional Grids\guilf{},
\newblock \Inname{} \textit{Proc. of the 10th Intl. Symp. on High Performance
  Distributed Computing (HPDC)}, \Aug{} 2001,  \pagesname{} 55-63.

\bibitem[FUR~02]{FurMayGouNewFieDar2002JPC}
\textsc{Furmento N.}, \textsc{Mayer A.}, \textsc{McGough S.}, \textsc{Newhouse
  S.}, \textsc{Field T.}\andname{}\textsc{Darlington J.}, \guilo{}{ICENI}:
  Optimisation of Component Applications within a Grid Environment\guilf{},
\newblock \textit{Journal of Parallel Computing}, \volumename\ 28, \numbername\
  12, 2002,  \pagesname{} 1753-1772.

\bibitem[Glo]{GlobusWeb}
The Globus Alliance: \url{http://www.globus.org/}.

\bibitem[KIC~04]{Sekitei}
\textsc{Kichkaylo T.}\andname{}\textsc{Karamcheti V.}, \guilo{}Optimal
  Resource-Aware Deployment Planning for Component-based Distributed
  Applications\guilf{},
\newblock \Inname{} \textit{Proc. of the 13th International Symp. on High
  Performance Distributed Computing (HPDC)}, Honolulu, HI, \Jun{} 2004.

\bibitem[LAC~04a]{LacPerPri2004CD}
\textsc{Lacour S.}, \textsc{Pérez C.}\andname{}\textsc{Priol T.},
  \guilo{}Deploying {CORBA} Components on a Computational Grid: General
  Principles and Early Experiments Using the {G}lobus {T}oolkit\guilf{},
\newblock \Inname{} \textit{Proc. of the 2nd Intl. Conf. on Component
  Deployment}, \volumename{} 3083 \ofname{} \textit{LNCS}, 2004,  \pagesname{}
  35-49.

\bibitem[LAC~04b]{LacPerPri2004RR5221}
\textsc{Lacour S.}, \textsc{Pérez C.}\andname{}\textsc{Priol T.}, \guilo{}A
  Network Topology Description Model for Grid Application Deployment\guilf{},
\newblock Research Report \numbername RR-5221, \Jun{} 2004, INRIA, IRISA,
  Rennes, France,
\newblock available at \url{http://www.inria.fr/rrrt/rr-5221.html}.

\bibitem[MDA]{MDA-deployment}
Deployment and Configuration of Component-based Distributed Applications
  Specification: \url{http://www.omg.org/cgi-bin/doc?ptc/2003-07-02}.

\bibitem[{Obj}~02]{CCM}
\textsc{{Object Management Group (OMG)}}, \guilo{}{CORBA} Components, Version
  3\guilf{},
\newblock Document \numbername formal/02-06-65, \Jun{} 2002.

\bibitem[PéR~03]{PerPriRib03IJHPCA}
\textsc{Pérez C.}, \textsc{Priol T.}\andname{}\textsc{Ribes A.}, \guilo{}A
  Parallel {CORBA} Component Model for Numerical Code Coupling\guilf{},
\newblock \textit{The International Journal of High Performance Computing
  Applications (IJHPCA)}, \volumename\ 17, \numbername\ 4, 2003,  \pagesname{}
  417-429, SAGE Publications.

\bibitem[WAN~03]{WanRodGil2003}
\textsc{Wang N.}, \textsc{Rodrigues C.}\andname{}\textsc{Gill C.}, \guilo{}A
  {QoS}-aware {CORBA} Component Model for Distributed, Real-time, and Embedded
  System Development\guilf{},
\newblock \Inname{} \textit{OMG Workshop On Embedded \& Real-Time Distributed
  Object Systems}, Washington D.C., \Jul{} 2003.

\end{thebibliography}

\end{document}